# THz optical beat-note detection with a fast Hot Electron Bolometer operating up to 31 GHz

G. Torrioli[1], A. Forrer[2], M. Beck[2], P. Carelli[1], F. Chiarello[1], J. Faist[2], A. Gaggero[1], E. Giovine[1], F. Martini[1], U. Senica[2], R. Leoni[1], G. Scalari[2] and S. Cibella[1*]

1 Istituto di Fotonica e Nanotecnologie, Via Del Fosso del Cavaliere 100, 00133 Roma, Italy.
2 Institute for Quantum Electronics, Department of Physics, ETH Zürich, 8093 Zürich, Switzerland
*sara.cibella@ifn.cnr.it

**Abstract:** We study the performance of an hot-electron bolometer (HEB) operating at THz frequencies based on superconducting niobium nitride films. We report on the voltage response of the detector over a large optical bandwidth carried out with different THz sources. We show that the impulse response of the fully packaged HEB at 7.5 K has a 3 dB cut-off around 2 GHz. Remarkably, detection capability is still observed above 30 GHz in an heterodyne beating experiment using a THz quantum cascade laser frequency comb. Additionally, the HEB sensitivity has been evaluated and an optical noise equivalent power NEP of 0.8 pW/$\sqrt{Hz}$ has been measured at 1 MHz.

## 1. Introduction

Hot Electron Bolometers (HEBs) are extremely sensitive detectors at frequencies above 1 THz and they are vastly employed in astronomical observations to map THz lines with a very high spectral resolution [1]. At the same time such devices present an intrinsically wide electrical bandwidth, making them appealing for applications in the THz range where the quest for high-speed detectors [2] is boosted by renewed interest in the field of ultrafast THz physics [3] and in the new studies of frequency comb technology [4,5]. For example, a new spectroscopic technique has been recently developed to demonstrate that all modes in the spectrum of a comb are uniformly spaced and phase coherent (i.e., SWIFT spectroscopy) [6]. Such measurement relies on a fast detector capable to detect the optical beat notes, typically lying in the 1-30 GHz regime. We investigate the performance of our detector in this scenario of high power THz sources [7] and extremely fast signals. Superconducting hot electron bolometers are considered an excellent technological solution because of their extreme sensitivity and response beyond tens of GHz [8]. The great advantage of this bolometer, made of an ultrathin layer of NbN film, is that it combines a small heat capacity with high heat conductivity resulting in a time constant down to 40 ps for small optimized devices [8]. Furthermore, compared to competing technologies like superconductor-insulator-superconductor tunnel junctions and Schottky diodes, the HEB does not possess any optical bandwidth limitation of the detection mechanism [9] and, as it is a fast thermal detector, it is useful throughout the IR and THz spectral regions [10,11].

Here we propose a phonon-cooled hot-electron bolometer coupled to a logarithmic spiral antenna [12] embedded in a coplanar waveguide that allows a packaging with a microwave end-launch connector. The aim of this paper is to demonstrate experimentally the high-speed operation of fabricated HEB using impulse response measurements [13]; it employs short (ps), broadband THz pulses emitted using a compact, commercial THz-TDS spectrometer as well as higher intensity signal produced by a quantum cascade THz laser frequency comb. The frequency response of the bolometer will be characterized using a wide bandwidth (>30 GHz)

RF spectrum analyzer. Furthermore, we provide a full characterization of our device on the basis of a hot electron model.

## 2. HEB fabrication and experimental set up

The HEB consists of a 1 µm-wide, 300 nm long and 6 nm thick NbN bridge fabricated on a highly resistive Si substrate. The superconducting layer has been deposited by means of a reactive dc magnetron sputtering in Ar/N2 gas mixture at 750C [10]. The bridge is connected to a self-complementary spiral antenna made of a 180 nm thick Ti/Au bilayer, embedded in a coplanar wave guide (with an impedance of ~50 Ohm). The inner diameter of the antenna is 7 µm. Starting from a silicon wafer coated with the NbN film, the HEB fabrication process uses the high resolution and realignment capability of a direct-writing electron beam lithography (EBL). In the first step, the Ti/Au layer is deposited on top of a PMMA resist mask, from which spiral antenna, wiring, contact pads and alignment marks are obtained by lift off. In the second step, we define the HEB bridge by using a mask of hydrogen silsequioxane (HSQ) resist and successively etching away the unprotected NbN film by reactive ion etching (see Fig 1).

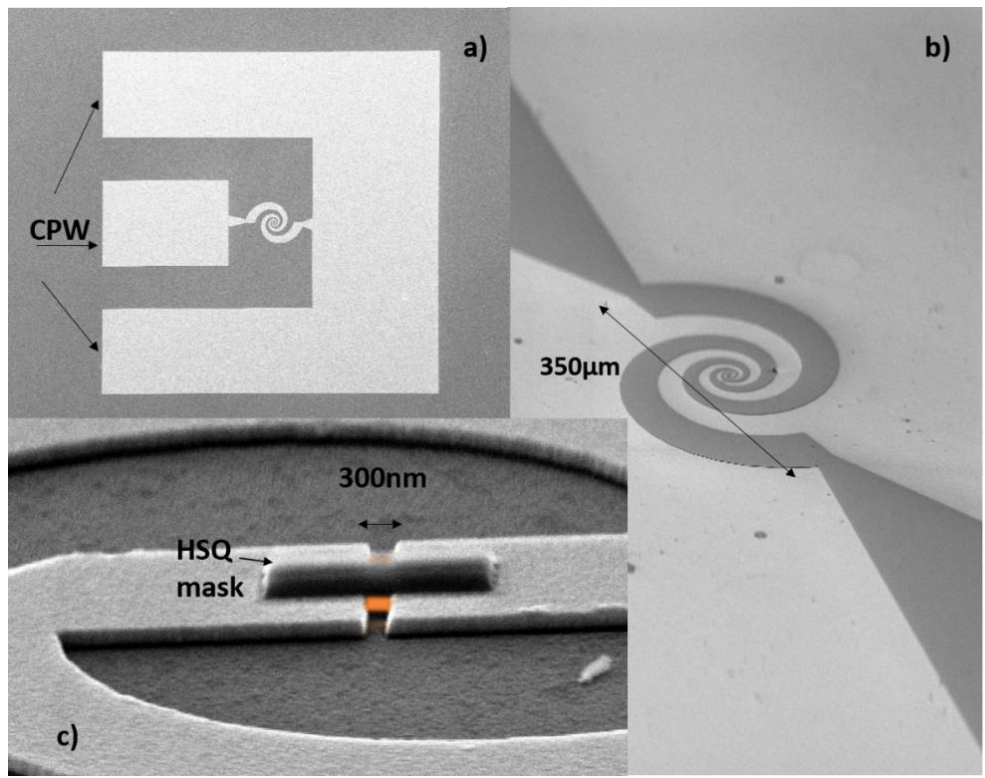

Fig 1. SEM micrograph of the device fabricated: (a) detail of the coplanar waveguide, (b) the spiral antenna and (c) the bolometric element at its centre (in the centre the detail of the HSQ mask and underneath the NbN layer (coloured in orange)).

The radiation coupling to the antenna is further enhanced by tightly pressing a high-resistivity hyper-hemispherical Si lens, 6 mm in diameter, onto the silicon backside surface. The fabricated HEB, is mounted onto a copper holder in vacuum, screwed on the cold plate and cooled down by using either: i) a helium flow cryostat for the heterodyne frequency measurements performed with the TDS system and the QCL source; ii) a pulse tube refrigerator for the voltage responsivity and NEP measurements. The optical windows of both cryostats are equipped with cold optical filters to suppress undesired wavelengths. In the set up utilized for the NEP measurements, we used a low noise cryogenic amplifier (LNA, CITLF1 by Cosmic Microwave Technology Inc 40 dB gain, 0.001–1.5 GHz bandwidth and noise temperature ~9 K) connected by means of a bias-tee to our HEB. To calibrate the HEB at 1 MHz we use a homogeneous multimode THz quantum cascade laser operating at 3 THz central frequency [14].

## 3. Impulse and frequency response

The frequency response of the HEB is studied by using optical short pulses generated by a mode locked Erbium telecom fibre laser and a photoconductive antenna of a commercial, time domain THz spectrometer (TeraSmart, Menlo Systems), combined with a spectrum analyser (SA) (see Fig. 2). In this case a cold, low-pass filter is mounted on the radiation shield of the helium flow cryostat, to cut off the infrared background.

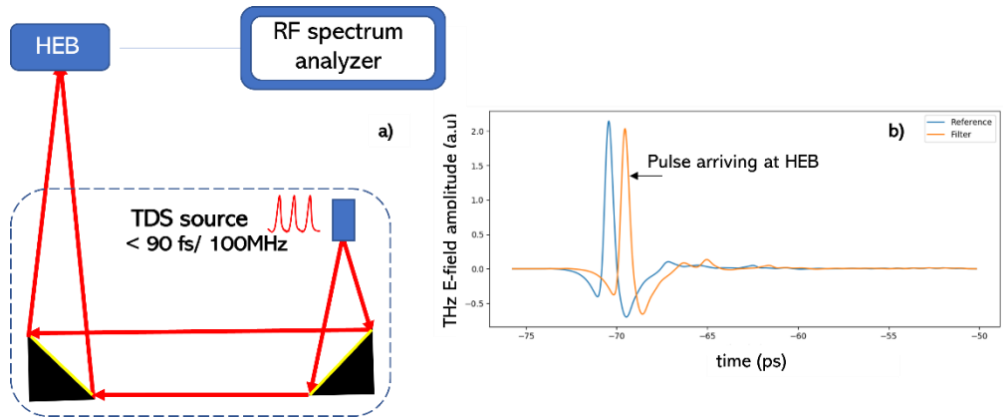

Fig 2: (a): Experimental set-up for measuring the impulse response of the HEB using THz picosecond pulses produced by a commercial photoconductive antenna illuminated by an Erbium-doped fiber mode-locked laser . (b): THz pulse arriving at HEB before and after the low pass filter used.

The THz pulse, induced by a femtosecond laser (<90 fs) with a repetition rate of 100 MHz, has a FWHM of about 0.5 ps and can be regarded as a good approximation a delta-function pulse with respect to GHz frequencies. For an ideally fast detector, i.e. flat frequency response, this THz signal would induce beating signals with nearly equal peak amplitudes within the GHz range we consider. Before going into the TDS results, we investigate the HEB I-V in the dark. Fig 3 shows the experimental I-V characteristics of the HEB, obtained with the TDS blanked but operating. We set the operating temperature for our measurements using a resistive heater and we took the family of I-V characteristics changing the bath temperature from 4.5K to 8.5K. The HEB fabricated has a normal state resistance (Rn) of 80 Ω and a critical current of 80 μA at 4.2K without illumination from THz-TDS.

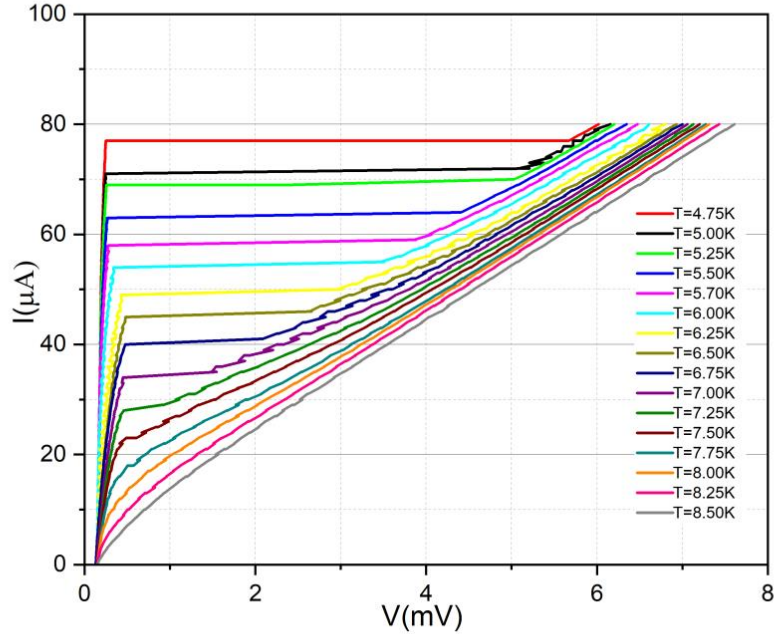

Fig 3: (a) Experimental current-voltage (I-V) characteristics of the HEB taken in different heating conditions (TDS signal blanked).

Now we unblock the optical path and measure for several temperatures and bias points the fundamental and the harmonic beating signals induced by the TDS system at frequencies (100, 200, 300,…., and so on up to 10000 MHz on the HEB with the SA). An example at 7.5K biased at 21uA is shown in fig. 4. The blue curve shows the raw trace from the SA , the red triangle mark the peaks of each beating signal and the green dots represents the corrected peaks by subtracting the RF path losses up to the wire bonds of the HEB. Therefore, the green curve in fig. 4, resembles to a good degree the intrinsic frequency response of the HEB. At frequencies above 10 GHz the residual path on the HEB chip should be further optimized since the RF wavelengths become shorter. The HEB is now biased at different currents of the I-V curves and its frequency response is measured as described before in the temperature range 4.5-8 K. When the HEB is operated at temperatures far from the Tc, by increasing the bias current, the I-V curves show a sudden jump from the superconducting branch to the resistive branch. On the other hand when the HEB is operated in a temperature range close to the Tc, the I-V curve jump does not occur anymore: this corresponds to a stable hotspot in the HEB which will expand gradually with the bias until the whole film goes into normal state [15]. In this temperature range the HEB can be polarized in each point of the I-V characteristics as long as dV/dI is continuous avoiding noisy instabilities. More in detail, we consider the behaviour of the HEB at 7.5 K (see fig 5). We mark the different bias points along the I-V with different colours (fig 5a). We first normalize all response curves to the maximum amplitude for all traces (in this case at 20μA, 100 MHz) highlighting the behaviour of the detector "gain". The bias point set at 20μA, i.e., the best in terms of gain, rolls off quite fast with frequency, while signal at 5 μA is nearly 30dB lower with a wider 3dB bandwidth (BW)  (see Fig. 5b). One should note that the 20uA bias point results in the highest gain at all measured frequencies despite the rapid frequency response drop. Then we normalize all intermode beating signals to the maximum for each curve which allows a better comparison of the 3dB BW. It can be clearly observed, that at this temperature the -3dB cut-off is around 2 GHz for lower detector gain and the largest band biasing the HEB at 5µA. This bias point would be optimal in application which need a flat response over a large BW. Therefore, combining the two graphs in fig. 5 we have all the information necessary to choose the optimal point where one would like to operate the HEB depending on the specific application.

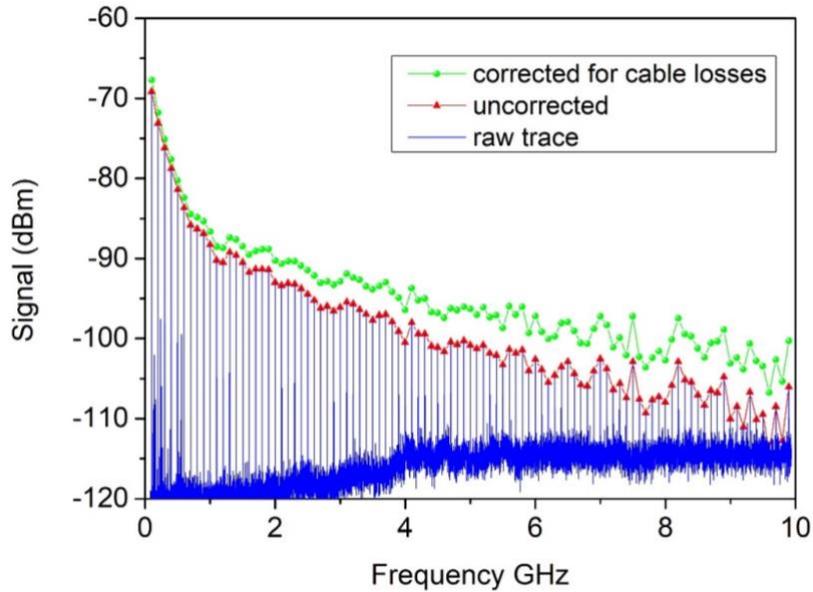

Fig 4: Harmonics of the erbium laser beat-note up to 10 GHz biasing the HEB at 21 µA at T=7.5 K. Data are corrected including losses due to RF cable in the cryostat.

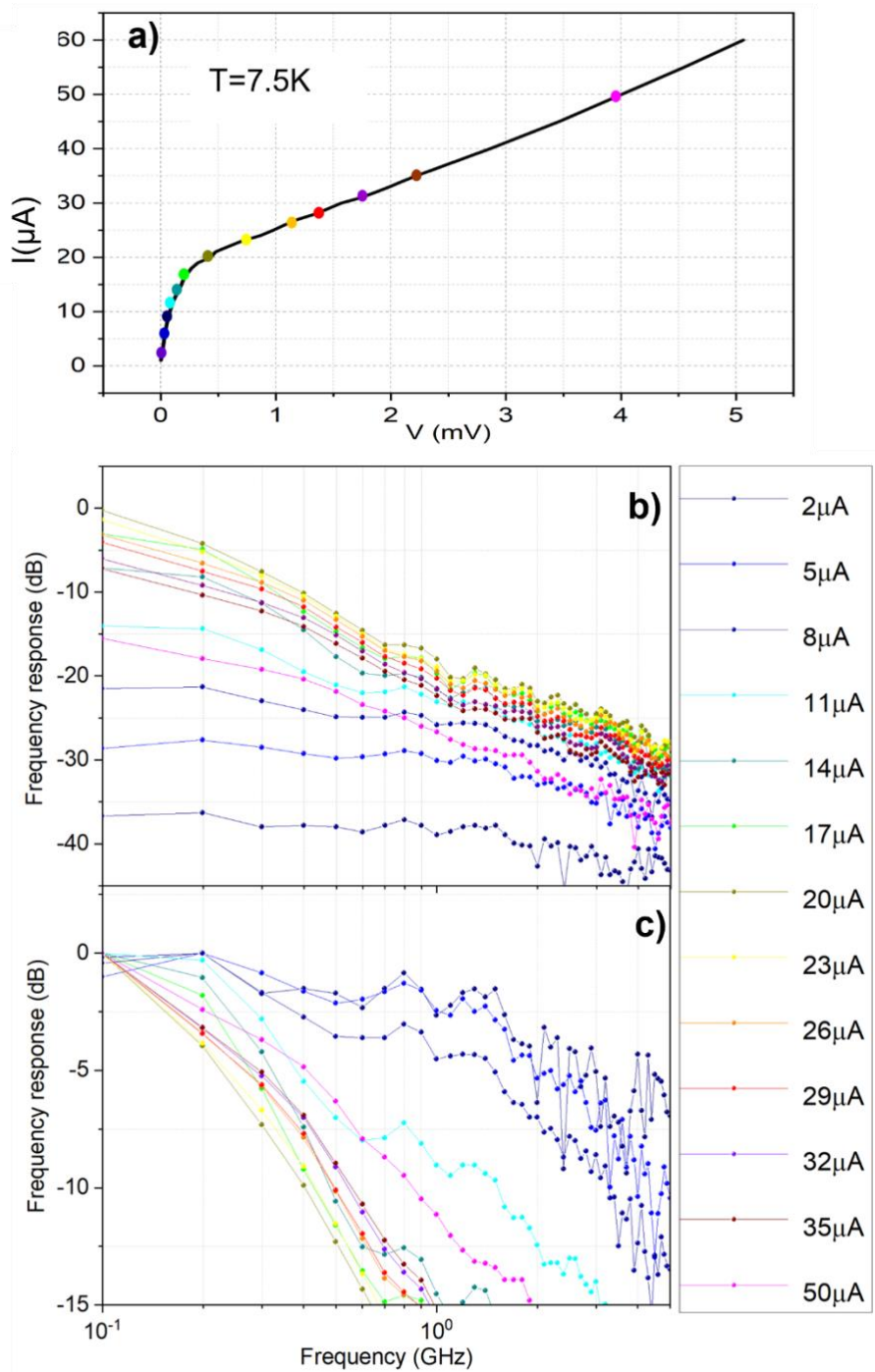

Fig 5: Frequency response of the HEB at T=7.5K at the different bias conditions. (a) I-V curve at 7.5K with coloured different bias points. (b) all response curves are normalized to the maximum amplitude for all traces (in this case at 20uA, 100 MHz). (c) normalizing all intermode beating signals to the maximum for each curve.

The same measurement data shown in fig. 5b at 7.5K can be further visualized in a 3D color map as shown in fig. 6. Here we did not normalize to the peak intensity to show the relatively high signal levels of up to 67 dBm of the beating signals, i.e. > 50 dB SNR. The white dashed lines represent the -3 dB and -15 dB cut off frequency.

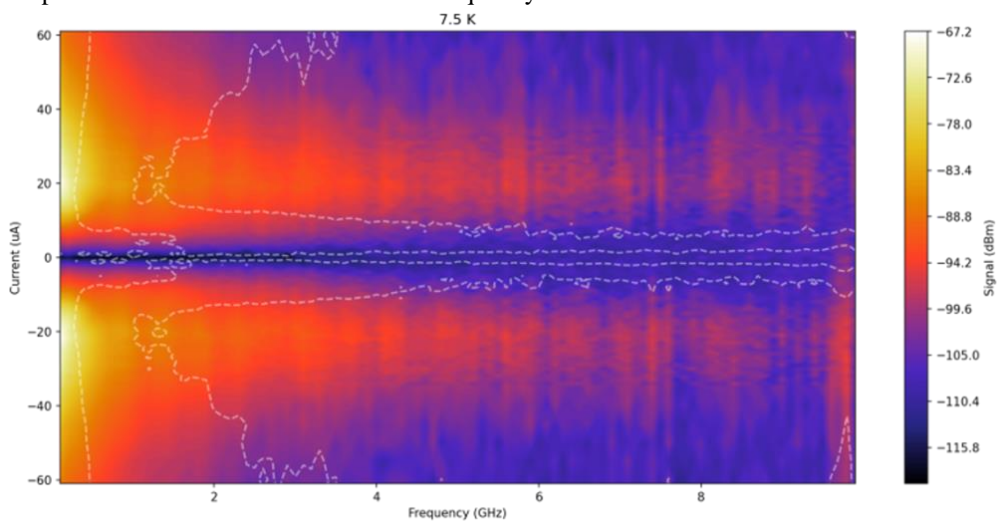

Fig 6. 3D colour plot: the amplitude of the signal taken at different bias points vs the frequency response and the bias current. The dashed line shows the -3 dB and -15 dB cut off. Changing the bias point along the I-V curve, the -3 dB and -15 dB cut off change.

Similar plots are presented for different temperatures in the supplement.

It is important to point out that the HEB is still sensitive above 10 GHz, but the THz signal produced by the photoconductive antenna is too low to be detected. To probe the ultimate frequency limit of the HEB we detect the intermodal optical beat note by an injection-locked THz QCL frequency comb, centered around 2.6 THz (see spectrum in Fig. 7(a)), with emission powers in the lower mW range. This device has an optical mode spacing 10.5129 GHz. It should be noted that this comb is in a mixture of frequency and amplitude modulated comb state, not in a pulsed mode locked state, and consists of relatively few number of modes compared to the TDS. Therefore, the beating signals are expected to differ in peaks power and we cannot extract the frequency response from this measurement reliably. Nevertheless we are able to detect the optical beat-note of the laser up to its third harmonic (31.5387 GHz signal, see fig. 7(b)), with a signal-to-noise ratio of 20 dB. It is important to note the analogy between Fig. 4 and Fig. 7(b), both showing the coherence of the combs even though in case of the THz QCL comb the electric field does not show pulses in the time-domain [5]. It is important as well to underline the fact that in the case of the THz QCL beat-note measurement no RF amplifier was employed because the signal extracted from the HEB cryostat was directly fed into the spectrum analyzer.

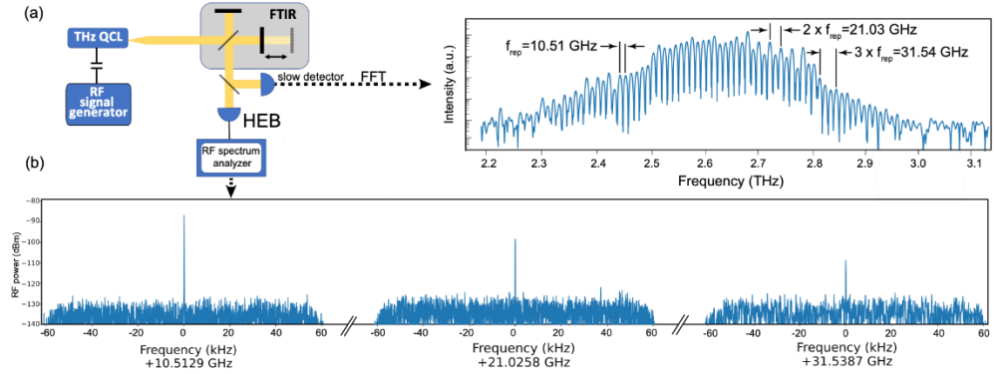

Fig 7 (a): experimental setup: an injection-locked THz QCL is coupled to a FTIR. The displayed spectrum corresponds to an 800 GHz bandwidth comb state with a repetition rate $f_{rep}$=10.15 GHz. (b): optical beat-notes detected with the HEB. Signals up to the third harmonic, corresponding to a frequency of 3x$f_{rep}$=31.54 GHz are detected with a SNR larger than 20 dB.

## 4. Device modeling

We used the two temperatures (2T) theoretical model [16, 17] to describe the electro-heating process that occurs in a HEB. In this way, we can extract the response of the HEB alone from the rest of the circuit in order to understand the measurements described in the previous sections. This model works for non-equilibrium superconductors maintained at temperature $T$ near the superconducting transition temperature $T$c. The classic description of the HEB under operation is that the incoming radiation and the electrical bias heat up the electrons in the NbN inducing a local resistance [18, 19]: an uniform enhancement of the electron temperature is assumed higher than the phonon-temperature. The response of the detector is controlled by the electron temperature, i.e., hot electrons, and by the local resistivity. Under operating conditions, the whole detector, including the contacts, is in the superconducting state, except for the centre of the NbN, where the resistance emerges due to the increase in the electron temperature. If the modulation of the electron temperature can follow the modulation due to the incoming signal also the resistivity of the HEB will be modulated and consequently a voltage signal is measured. Following the 2T model the superconducting NbN film can be described in terms of two coexisting subsystems: electrons and phonons. If the temperature T is close to the critical temperature Tc the effective temperatures of the electron and phonon subsystems (Te and Tph, respectively) are assumed to be established instantly and uniformly. With this hypothesis the hot electron effect in superconductors can be described by [20]:

$$\frac{dT_e}{dt} = -\frac{T_e - T_{ph}}{\tau_{e-ph}} - \frac{T_e - T_b}{\tau_{diff}} + \frac{j^2 \rho(T_e, I_b) + p(t)}{c_e} \qquad (1)$$

$$\frac{dT_{ph}}{dt} = \frac{c_e}{c_{ph}} \frac{T_e - T_{ph}}{\tau_{e-ph}} - \frac{T_{ph} - T_b}{\tau_{esc}} \qquad (2)$$

where, $T_b$ is the bath temperature, $\rho(T_e, I_b)$ is the local resistivity of the HEB in the normal state, j is the current density and p(t) is the rf power absorbed per unit volume. The electron-phonon interaction time and phonon escape time are $\tau_{e\text{-ph}}=500\,T_e^{-1.6}\,psK^{1.6}$ [15, 20] and

$\tau_{esc}= 10.9 \cdot d$ ps nm$^{-1}$, respectively. The film thickness is d, the diffusion coefficient in the NbN is D=0.45cm$^{-2}$ s$^{-1}$ L the bridge length and $\tau_{diff}=L^2/\pi^2 D$ is the effective diffusion time constant. The electron and phonon specific heats are defined as $c_e=1.85\times10^{-4}T_e$ Jcm$^{-3}$ K$^{-2}$ and

$c_{ph}=9.7\times10^{-6}\,T_e^3$ Jcm$^{-3}$ K$^{-4}$. In an HEB based on an ultra-thin NbN film deposited on a silicon substrate the relaxation times are typically $\tau_{e\text{-ph}} \approx$ 10-15 ps and $\tau_{esc} \approx$ 50 ps$\ll\tau_{diff}$ (phonon cooling) [15]. To predict the current to voltage characteristics with the 2T model, we start from the measured transition of the resistance (current-dependent) vs T of our HEB. Following the model proposed by Hajenius et al. [19] we approximated the R(T) curve to a Fermi-like (sigmoidal) function as introduced by Gershenzon et al [21]:

$$\frac{\rho(T_e)}{\rho_n} = \frac{1}{1+e^{-(T_e-T_c)/\Delta T_c}} \quad (3)$$

which represents the local resistivity for a given local electron temperature $T_e$ with $\rho_n$ the resistivity in the normal state and $T_c$ the critical temperature of our bolometer. The intrinsic transition is measured for different dc bias current. Fig. 8 shows the experimental points (symbols) and the modelled curves (continuous lines).

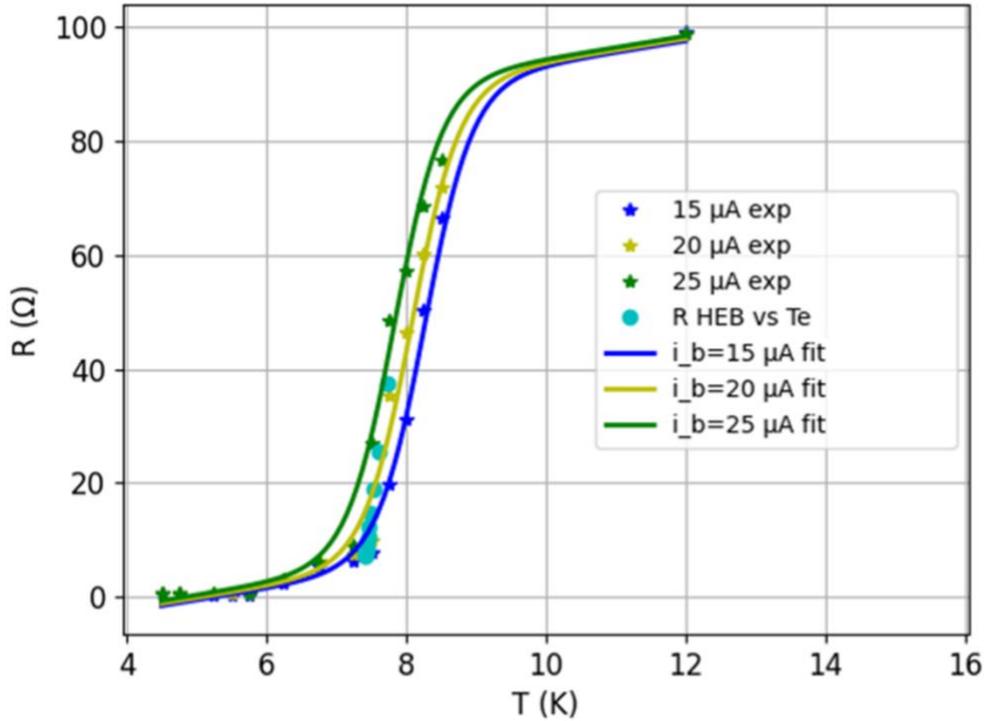

Fig 8: The intrinsic transition measured for different dc bias currents ranging between 15 and 25 µA and the modelled curve following the sigmoidal fit. We also graph the RHEB vs Te for different $I_b$

Increasing the current, the R (T, $I_b$) curves shifts to lower temperatures [18] and the downshift of the critical temperature Tc follows the equation

$$\frac{I_b}{I_c} = \left(1 - \frac{T_c(I_b)}{T_c(0)}\right)^\gamma \quad (4)$$

The ρ(T, $I_b$) resulting from R (T, $I_b$) is used as HEB fingerprint in eq 1 (2T model) and now with the 2T model is possible to predict the complete I-V characteristics.

Fig 9 shows the experimental I-V curves at 7.5 K taken with the TDS on and TDS off. These curves are in good agreement with a set of I-V points calculated according to the 2T model, in the limit $T_e \leq T_c(I_b)$, for a temperature T = 7.25 K and an impinging radiation power respectively of 1 pW and 11 nW. Starting from this result we can calculate the frequency response for different bias currents of the detector.

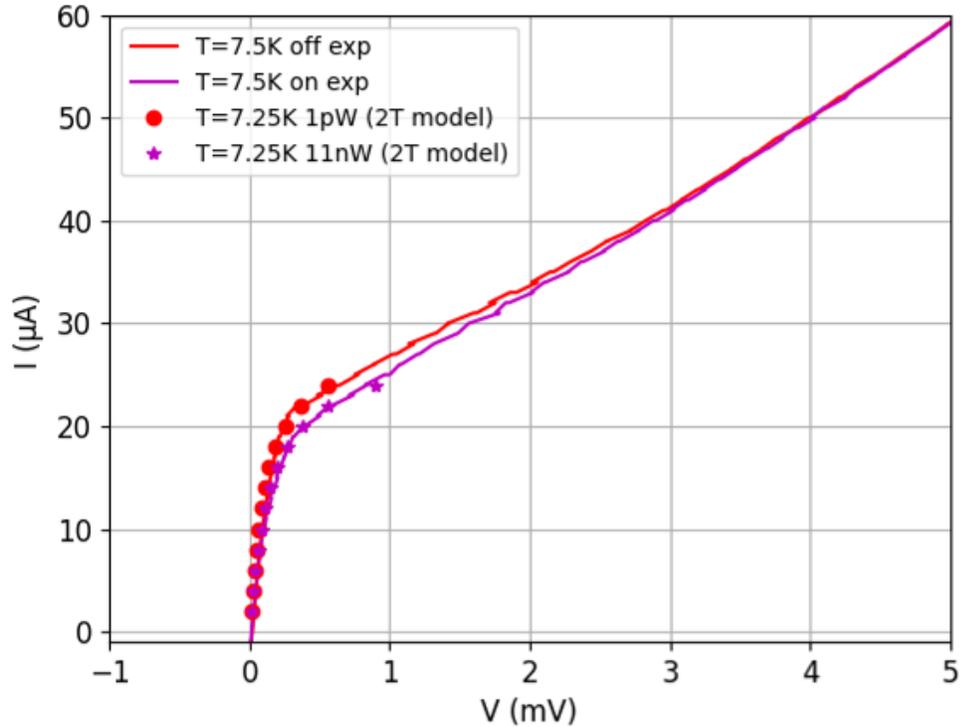

Fig 9: Experimental I-V at 7.5K with laser off (red continuous line) and laser on (magenta, continuous line) with superposed the results of the 2T model at T=7.25K with power P=1pW (red dots) and P=11nW (magenta stars).

Fig. 10 shows the simulated frequency response of the HEB at 7.25K at two different bias points 10 and 20 µA showing, independently from the $I_b$, a 3dB frequency cutoff ≥ 2GHz. On the other side, the high frequency cutoff of the experimental response has a strong dependence from the $I_b$ (0.6GHz and 0.15GHz for 10 and 20 µA, respectively). The experimental high frequency cut off is compatible with a first-order low pass filter cut off frequency, probably due to parasitics in the circuit (wire bonds, cryo-mount and cables) and has a strong dependence with the $I_b$: changing the operating point of the HEB the impedance matching is changed accordingly [22, 24, 27].

In addition, in the flat region (around 100 MHz) of the experimental data, we have a gain reduction going from 20 to 10 µA, which is in agreement with the reduction resulting from the simulation.

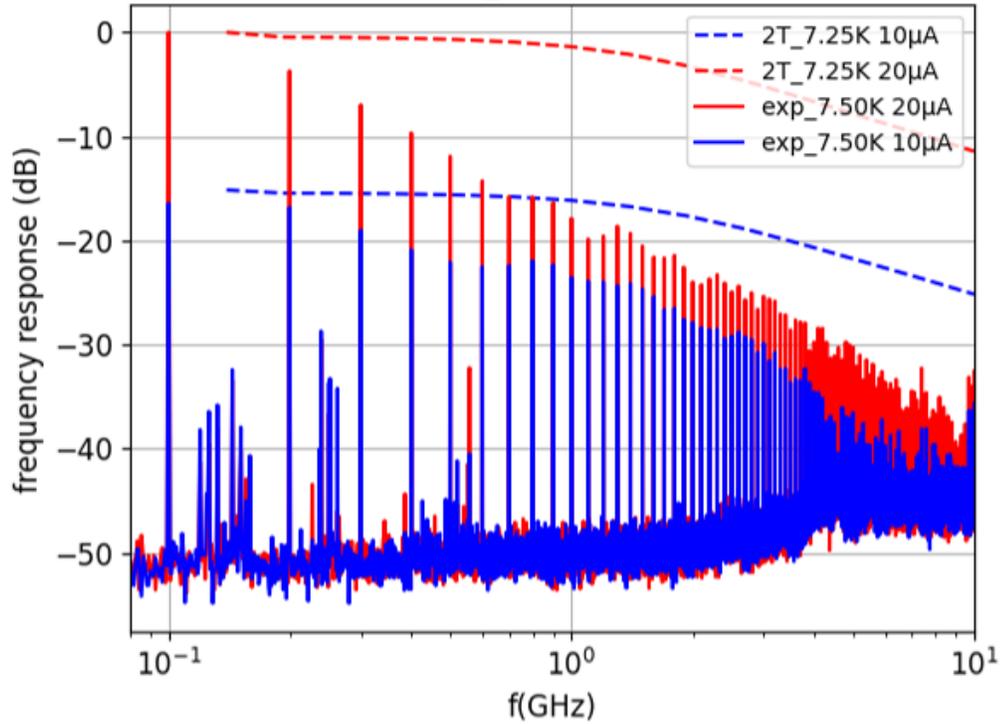

Fig 10. Frequency response of a HEB at 7.5K at Ib=10µA (blue continuous line) and Ib=20µA (red continuous line) and the result of the 2T model at 7.25K (dashed lines).

Figure 11 (a, b) shows the experimental data with the simulated frequency response including an additional first-order low pass filter, in the two cases. Applying this procedure to other bias currents $I_b$, we obtain the result shown in fig 11 c) where the corresponding cut-offs are displayed

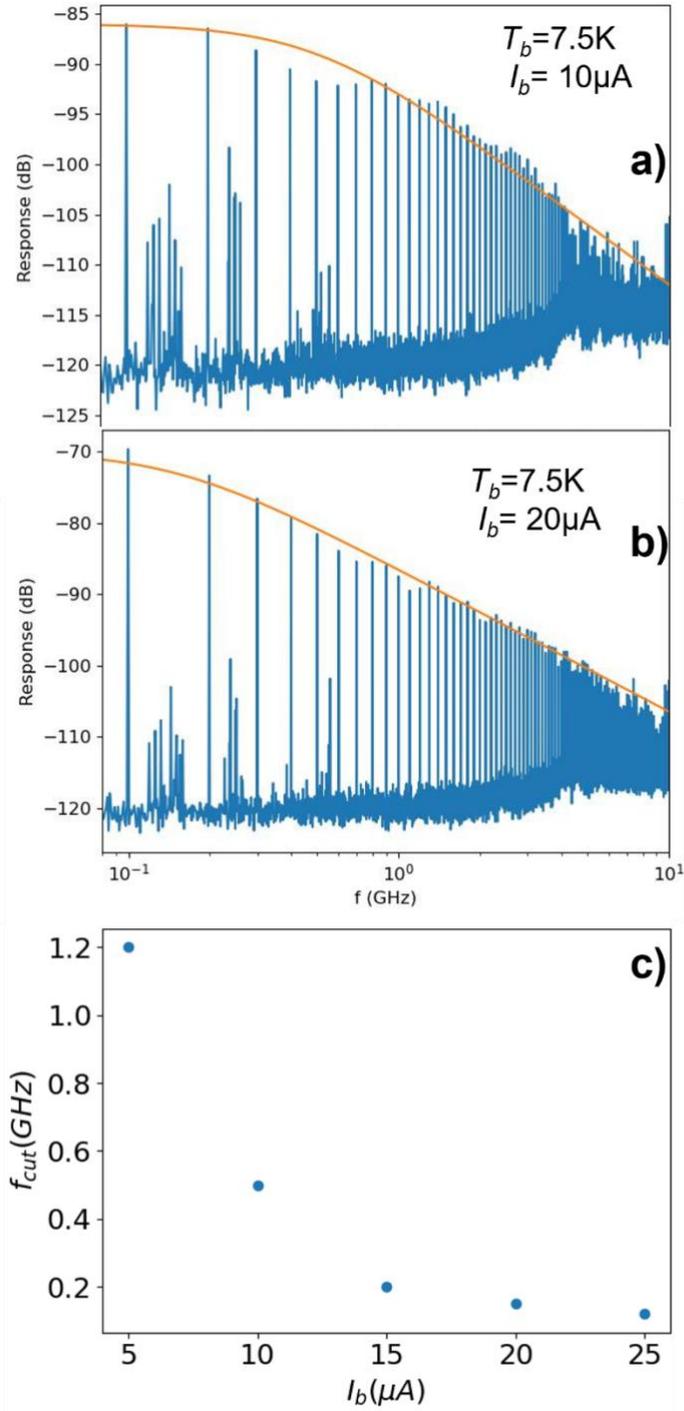

Fig 11: Simulated frequency response of the HEB introducing a low pass filter. a) result for T=7.5K Ib=10µA; b) T=7.5K, Ib=15µA, c) frequency cutoff vs Ib.

## 5. Responsivity and NEP

Important figures of merit to define HEB performances are the responsivity and its sensitivity [23], the latter characterized by the noise equivalent power (NEP), i.e., the minimum incident power required to obtain a unit signal to noise ratio in a 1Hz bandwidth. To calibrate the HEB we used a high-power THz-QCL as input source with the active region based on four-quantum well super-diagonal structure [14] operating at 77K mounted in a small nitrogen cryostat (about 2dm$^3$ in volume). To map the intensity of the beam pattern of the QCL and perform power measurements calibration [1] for this specific QCL we used a room temperature commercial pyro-electric detector (GENTEC model QS2-THz-BL) with an aperture diameter of 2 mm, responsivity 140kV/W and a black HDPE window to filter the IR.

We performed a 2D map of the radiation pattern of the QCL working in pulsed mode with a duty cycle of 5%. By integrating the intensity of the beam patterns we got an output power of about 20 µW, which corresponds to a total power in continuous wave of 0.4 mW.

We now perform a new map by using the HEB operating the QCL in the same condition. If we integrate the intensity for all pixels and take in account the cross section between the single pixel area of pyro-electric detector and the spatial resolution of the HEB, we got a responsivity of 560 V/W. We can now infer the NEP of our HEBs.

We operate the detector in direct mode setting the operating temperature with the resistive heater. The HEB can be polarized in each point of the I-V characteristics as long as dV/dI is continuous to avoid noisy instabilities in the HEB, as we said in the previous section. The detector NEP, is obtained by the ratio between the voltage noise spectral density Vn(f) measured at the HEB output and the HEB responsivity Res(f). To measure the voltage noise density, we used the cryogenic amplifier and a room temperature amplifier.

We point out that in these NEP measurements our setup is limited in the MHz range. Changing the bath temperature and the current bias we obtained a NEP below 1pW/$\sqrt{Hz}$ biasing the device between 15 and 30 µA at 7.5 K with a minimum NEP = 0.8 pW/$\sqrt{Hz}$.

We believe that we are underestimating the HEB sensitivity due to the direct detection effect [25], which shifts the optimal operating point to a higher voltage when using a calibration load higher than 100nW.

We cannot provide the NEP value in the GHz band but we can estimate an upper limit in this frequency range, considering the responsivity reduction due to the attenuation that the signal has at higher frequencies (see fig. 4) and assuming, as a worst case scenario, that noise is not affected by this reduction effect and remains flat.

## 6. Conclusions

By using two different sources, a picosecond THz laser-based source and a THz QCL frequency comb we have shown that our phonon-cooled HEB is still highly sensitive up to 30 GHz once operated in mixer mode detecting up to the third harmonic of the beating signal of a THz QCL comb with a round trip of 10 GHz. In addition, we measured the HEB sensitivity estimating a minimum NEP of 0.8 pW/$\sqrt{Hz}$, although the direct detection effect led to underestimate the optical responsivity. We provide a fully characterization of our detector in terms of frequency response, sensitivity and NEP. We show that the fabricated HEB has a 3dB around 2 GHz cutoff at 7.5 K. Accordingly to the theoretical model we show how the frequency response of the HEB changes, changing both the bias point and its operating temperature. Such detector has already been employed for SWIFTs characterization of THz QCL frequency combs [30]

Our next challenge is to optimize the HEB technology in terms of the rf coupling (coplanar waveguide design on chip as well as the rf packaging on the rf connector) but also from the bolometer technology point of view, i.e., improving the quality of the contact pads interface itself: all ingredients needed for targeting a higher frequency cut-off limit [27-29]. This high-speed and high-sensitivity detector can find several applications in combination with THz frequency combs for the diagnostic of the comb itself but also in the field of atomic and molecular broadband spectroscopy [31].

*Fundings*

Actphast 4 Researchers P2020-41
G.S. would like to acknowledge support from the ERC Grant CHIC (No. 724344)

*Acknowledgements*

S.C. and G.S. would like to thank Juraj Darmo for helpful discussion